\documentclass[letterpaper, 10 pt, conference]{ieeeconf}  

\IEEEoverridecommandlockouts                              

\usepackage{graphicx} 
\usepackage{epsfig} 
\usepackage{mathptmx} 
\usepackage{times} 
\usepackage{amsmath} 
\usepackage{amssymb}  
\usepackage{mathrsfs}  
\usepackage{url}

\usepackage{color}
\usepackage{xcolor}
\usepackage{caption}
\usepackage{preambles}

\usepackage{caption}
\usepackage{subcaption}
\title{\LARGE \bf
Adaptive Feedback Regulator for Powered Lower-Limb Exoskeleton under Model Uncertainty
}

\author{Kirtankumar J. Thakkar$^{1}$, Victor C. Paredes$^{1}$, and Ayonga Hereid$^{1}$

\thanks{This work is supported by Prof. Hereid's start-up package from the Ohio State University. The authors particularly thanks the engineers from Wandercraft for their consistent assistance in this project.}

\thanks{Mechanical and Aerospace Engineering, Ohio State University, Columbus, OH, USA
{\tt\small \{thakkar.111,paredescauna.1,hereid.1\}@osu.edu).}}
}

\begin{document}
\maketitle

\begin{abstract}
This paper presents a neural network (NN) based adaptive feedback regulator to ensure the lateral and longitudinal stability and regulate the desired walking velocity of a lower-limb exoskeleton under model uncertainty. The traditional model-based controllers for lower-limb exoskeletons often fail to stabilize the robot or accurately track the desired behaviors under model uncertainties or external disturbances. This paper proposes a neural network (NN) based online adaptive regulator that compensates for the unknown changes in model parameters and external disturbances by modifying the nominal joint trajectory. A gradient descent-based delta rule is implemented to update the weights of a single layer NN, which can be efficiently performed online by design. We demonstrate the performance of the presented regulator on ATALANTE, a fully actuated lower limb exoskeleton designed for paraplegic patients. The simulation results show that the proposed approach noticeably improves stability and the tracking performance of the system, despite significant changes in model parameters and large adversarial pushes. 
\end{abstract}



\section{Introduction}


Robotic lower-limb exoskeletons can be used to assist individuals with walking disabilities to perform normal ambulatory functions and to provide exceptional health benefits, as well as positive psychological effects~(\cite{raab2016effects,geigle2017exoskeleton, krebs1998robot}.)
Many lower-limb exoskeletons heavily rely on arm-crutches and hand-tuned hierarchical control algorithms to provide supports to the user, resulting in inefficient walking and limiting the user from utilizing their upper limbs in other purposes and potentially leading to discomfort~(\cite{jimenezfabian2012review, anam2012active}.) In recent years, a growing number of successful studies translate the formal model-based controllers designed for bipedal robots to powered lower-limb exoskeletons, realizing hand-free dynamic walking without the use of crutches (see \cite{agrawal2017first,harib2018feedback,gurriet2018towards, duburcq2020online,tucker2020preference}.) By capturing the underlying dynamics more accurately, model-based approaches yield more natural dynamic walking behaviors. Nevertheless, they are very susceptible to changes in the system's kinematic and dynamic properties and external disturbances. To realize an expected behavior, such as walking at a certain desired velocity, one needs to model the human-exoskeleton system as accurately as possible and generate an appropriate gait trajectory---typically via optimization---for this specific model. However, once the circumstance changes (e.g., the user holds heavy objects, different users, etc.), the gait generation process must be performed again to ensure the expected performance. 
For instance, \cite{Chevallereau2003RABBIT} showed significant velocity tracking errors when mass and inertia parameters changing by $\pm 20\%$.
It is also important to note that the gait generation process is typically not fast enough to perform online due to the complex nonlinear system dynamics. 

Recent studies leverage bio-inspired feedback regulators to improve the robustness of model-based controllers for bipedal robots. For instance, \cite{Gnucci2019OnTheAdaptive} uses the non-collocated adaptive virtual control to achieve sustained locomotion for simple bipedal robots. \cite{Aoustin2003Control} shows that the stability of bipedal robots improves by regulating the step length of the gaits in response to pelvis velocity. \cite{Da2016From2D} and \cite{Gong2019Feedback} introduces swing foot placement regulators---which varies the swing foot landing position in response to the robot walking velocity---to stabilize 3D walking robots. While these heuristically-design regulators improve the robot's overall stability, they still yield noticeable steady-state tracking errors of desired behaviors (e.g., forward walking velocity) in the presence of significant model and environmental uncertainties---scenarios that wearable assistive devices often have to face. While the previous methods focused on heuristic regulators, \cite{castillo2020velocity} developed a model-free learning-based adaptive regulator for an underactuated 3D bipedal robot, Cassie. This approach improved the steady-state tracking performance by compensating the system's unknown dynamics with an adaptive learning scheme on the transient error dynamics of walking velocities. 

\begin{figure}
 \centering
 \vspace{2mm}
 \includegraphics[trim={0cm 0cm 0cm 0 cm},clip,width=1\columnwidth]{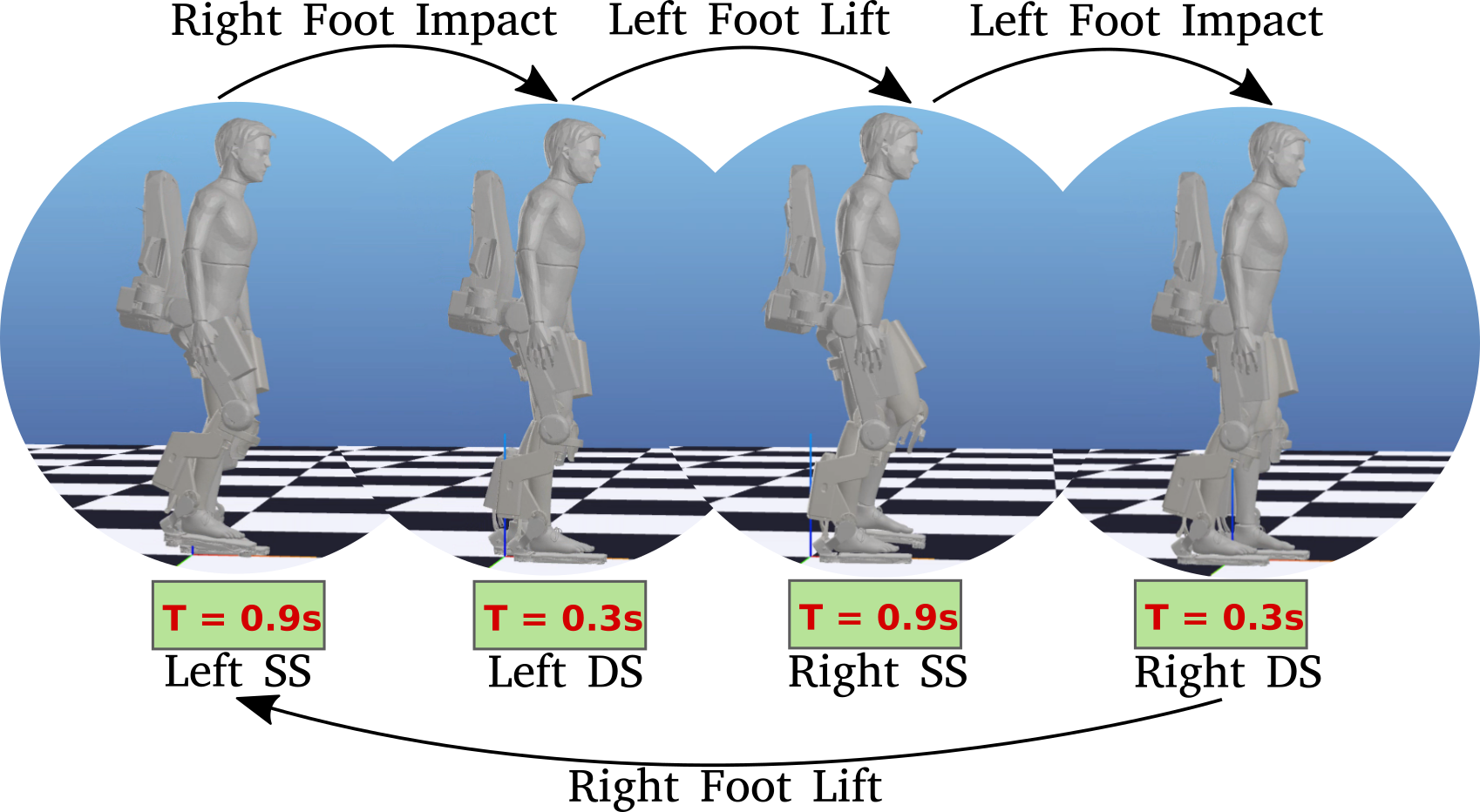}
 \caption{A directed graph representing four domains, left single support, double support, right single support, and double support, of the periodic flat-footed walking of ATALANTE.} 
 \label{fig:graph}
 \end{figure}

Build upon our previous work in \cite{castillo2020velocity}, this paper presents a neural network (NN) based adaptive feedback regulator for a fully actuated exoskeleton, named ATALANTE, to improve the stability and accurate tracking of desired walking velocities under model uncertainties. The whole process to realize stable walking starts from designing a nominal gait pattern via a model-based trajectory optimization. A nominal gait for lightweight biped robots generally comprises two single support domains. Keeping its heavyweight and stability in mind, we added two double support domains (see \figref{fig:graph}).  In this paper, we use the Matlab package FROST (\cite{frost}) to generate the nominal gaits for a range of forward walking velocities for a specific user. 
A model-free neural network-based adaptive feedback regulator is then developed to regulate the walking velocity of the exoskeleton under a wide range of model uncertainties and large external disturbances. 
The regulator can be implemented online without any prior knowledge of the mechanical properties, neither the exoskeleton nor the patient.
In particular, we developed a gradient descent delta rule to train the weights of a single layer neural network policy, making it directly transferable to robot hardware and evaluated in real-time without any modification. The primary input for the neural network is the horizontal pelvis velocity (see \secref{adapt_reg}) which is related to the overall stability since it is closest to the overall center of mass of the system. We are using the IMU sensor present on the pelvis link and leveraging the invariant extended Kalman filter (InEKF) (\cite{Ross2020EKF}) to estimate the pelvis velocity with respect to the world frame. 
The proposed approach shows consistent performance in simulation despite significant variation of the model properties of the system. 
In particular, we demonstrate its robustness for several different walking speeds of the robot for different masses, mass distributions, link geometries, and external impulsive disturbances. We further summarize the primary contributions of the present paper as follows:
\begin{itemize}
    \item Two double support domains have been added in the design of a nominal gait, leading to an overall improvement in the stability.
    \item Inspired from the regulator developed for the Cassie biped robot, we have developed and implemented a NN based adaptive feedback regulator for ATALANTE.
    \item In the end, we have presented a study between the proposed regulator and the heuristic foot placement regulator to highlight the overall performance improvement.
\end{itemize}

\begin{figure}
\centering
\vspace{2mm}
\includegraphics[trim={0cm 0cm 0cm 0cm},clip,width=0.9\columnwidth]{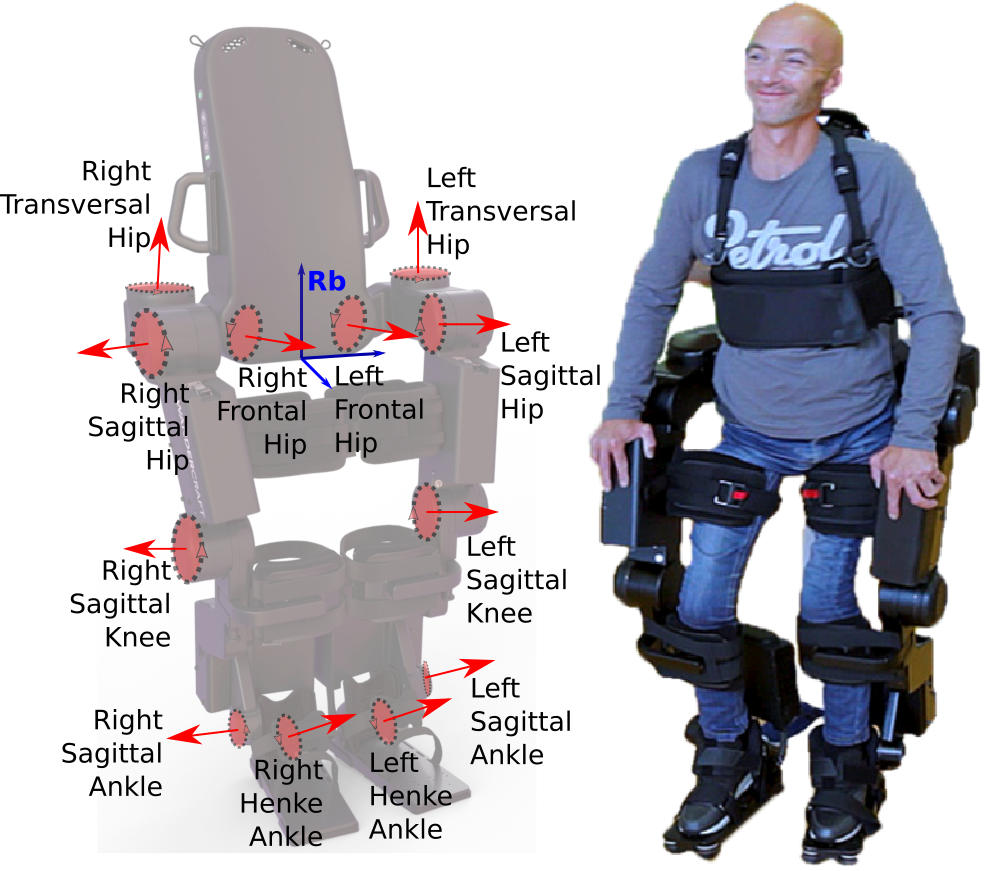}
\caption{The ATALANTE exoskeleton and its kinematic tree structure, where $R_b$ is the floating-base coordinate located at the center of the pelvis.} 
\label{fig:atlante}
\end{figure}


The remainder of the paper is organized as follows. \secref{section2} states the system model and gait generation process of the human-exoskeleton system, \secref{section3} presents our main contribution, the NN-based adaptive feedback regulator for robust locomotion of lower-limb exoskeletons under uncertainty. Finally, \secref{section4} shows the simulation results of the controller on ATALANTE, and concluding remarks are provided in \secref{sec:conclusion}.


\section{PROBLEM FORMULATION} \label{section2}


This section, presents the mathematical formulation of the hybrid zero dynamics (HZD) based walking controller for the lower-limb exoskeleton. We consider the system dynamics's hybrid nature as it comprises a continuous walking motion and discretely modeled ground impacts. We aim to generate a continuous periodic orbit of the robot joints and enforce them via virtual constraints to realize a stable walking motion.

\subsection{Human-Exoskeleton Lumped System Model}
\label{sec:robot-model}
ATALANTE is a fully-actuated lower-limb exoskeleton designed for improving the mobility of patients with paraplegia~\cite{harib2018feedback}. It consists of twelve actuated joints. The weight of the device is 75 kg and can fully support up to a 90 kg user (see \figref{fig:atlante}). To model the human-exoskeleton system, we consider the user body is rigidly attached to the exoskeleton to form a lumped mass system. The length of the thigh and shin links of ATALANTE can be adjusted to the user's body measurements. The mass, inertia, and center of mass of each link are approximated by the typical mass distribution of a human from its weight, assuming all links are rigid~\cite{winter2009biomechanics}. 
The generalized coordinates of the lumped system can be expressed by
\begin{align}
    \label{eq:coordinates}
    q  = (p, \phi, q_b) \in \mathbb{Q},
\end{align}
where, $p\in\R^3$ and $\phi\in SO(3)$ are the global position and orientation of the floating-base coordinates rigidly attached at the pelvis link, and $q_b \in \R^{12}$ is the vector of relative angles of all actuated joints, as shown in \figref{fig:atlante}. 
The walking motion of an exoskeleton is described as a hybrid control system  triggered by the changes in the ground-foot contact conditions. The hybrid control system $\mathscr{HC}$ can be described as:
\begin{align}
   \label{eq:hc}
   \mathscr{HC} = \{\Gamma, \mathscr{D}, \mathscr{U}, S, \Delta, FG\},
\end{align}
where, $\Gamma$ is the directed graph, $\mathscr{D} \subset T\mathbb{Q}$ is a smooth manifold representing the admissible domain of the system states, $\mathscr{U}$ represents the admissible controls, $S \subset \mathscr{D}$ is the switching surface, $\Delta$ is the reset map, and $FG$ represents the equations of motion of the continuous dynamics, described in \eqref{eq:eom}. With respect to previous work \cite{castillo2020velocity}, we considered a more human-like two-step periodic walking of the exoskeleton consisting of four domains, including two single domain phases and two non-instantaneous double support phases that constrains both feet to be in contact with the ground, as shown in ~\figref{fig:graph}. 
Given $u \in \R^{12}$ represents the actuator input vector and $\lambda$ is the contact wrench vector at the contact points, the continuous dynamics of the system is given as
\begin{align}
    \label{eq:eom}
    M(\q) \ddq + H(\q,\dq) = B u + J_d^T(\q) \lambda,
\end{align}
where $M(q)$ is the inertia matrix, $H(q,\dq)$ is the collection of Coriolis, centrifugal, and gravity forces, $B$ is the actuator distribution matrix, $J_d(q)$ is the contact Jacobian matrix for a domain $d\in \mathscr{D}$. During the simple support phases, the Jacobian $J_d$ maps the wrench from the unique contact foot. However, during double support phases, $J_d$ maps the action of the contact wrenches on both feet by appending the appropriate individual jacobians. The discrete dynamics of the system is determined by assuming all impacts are inelastic and occur instantaneously~\cite{Grizzle2014Models}. 
For more details regarding the hybrid system model's construction, we refer the readers to~\cite{Grizzle2014Models,hereid2018dynamic}.

\subsection{Gait Generation via Trajectory Optimization} 


Given the hybrid system defined above, we can formulate the gait generation problem as a nonlinear trajectory optimization problem using FROST. It is a Matlab based open-source software environment for developing a model-based control and motion planning for robotic systems whose dynamics is hybrid in nature. FROST formulates the trajectory optimization problem for such systems using \emph{direct collocation} methods (see \cite{frost}). Given the hybrid system model in \eqref{eq:hc}, the optimization problem can be defined as
\begin{align}
  \label{eq:gait-opt-frost}
  \hspace{-2px}\argmin  &\sum_{d \in\mathscr{D}}L_d(q,\dq,u)\\
  \hspace{-5px}\st &\tag{dynamics} M(\q) \ddq + H(\q,\dq) = B u + J_d^T(\q) \lambda, \\
                        &\tag{collocation} \delta_d(x_0,\dots,x_i,\dot{x}_i,\ddot{x}_i,\dots,\ddot{x}_{N}) = 0,\\
                        &\tag{reset map} (x^{d^+}_0,\dot{x}^{d^+}_0) = \Delta_{d}(x^{d}_{N},\dot{x}^{d}_{N}),\\
                        &\tag{path constraints} C_d(x_i,\dot{x}_i,u_i) \geq 0
\end{align}
where, $L_d(\cdot)$ represents the cost function defined in the domain $d\in\mathscr{D}$, $\delta_d$ is the collocation constraints, $\Delta_{d}$ is the reset map associated with domain $d$ and its subsequent domain $d^+$, and $C_d(\cdot)$ is the collection of physical constraints. 



In addition to necessary hardware constraints, such as joint motion and actuator torque limits, 
we employ several design-specific constraints to limit the solution space of the optimization problem to realize natural human-like walking patterns.
For instance, upright torso orientation is expected during human-like walking, which can be translated into a constraint as follows,
\begin{align}
    \phi_{min}^{tor} \leq \phi^{tor}(q) \leq \phi_{max}^{tor}.
\end{align}
The swing foot should achieve a minimum height from the ground during a step to avoid inadvertent and early ground impacts. Similarly, higher swing foot height can be uncomfortable for the user. This can be guaranteed by the following constraints
\begin{align}
    h_{min}^{sw}(q) \leq h^{sw}(q) \leq h_{max}^{sw}(q),
\end{align}
where, $h^{sw}(q)$ is the height of the swing foot, $h_{min}^{sw}(q)$ and $h_{max}^{sw}(q)$ are functions that profile the minimum and maximum foot clearance over a step, respectively.

Moreover, the swing foot must impact the ground with low negative velocity to produce a practical step. Note that a positive velocity would not provide a feasible solution in this context, and a large impact velocity can induce vibrations, even damages, on the robot. Such a constraint can be formulated as,
\begin{align}
    - \dot{h}_{max}^{sw} \leq \dot{h}^{sw}(q_N,\dot{q}_N) \leq 0,
\end{align}
where, $\dot{h}_{max}^{sw}$ is the maximum allowed impact velocity.
Further, the ground reaction wrenches $\lambda$, must enforce the contact wrench cone to avoid slipping and the ZMP condition to guarantee that the support foot remains in complete contact with the ground, i.e.,
\begin{align}
    \lambda \in \mathcal{B}_{cws},
\end{align}
where $\mathcal{B}_{cws}$ represents the contact wrench cone determined by the geometry of the foot contacts~\cite{dai2014whole}.

It is important to note that the model-based gait optimization problem is constructed for a specific user's dynamical model, considering the user's geometric and weight measurements.. Each gait is also designed to achieve a specific desired walking velocity by imposing an equality constraint in the optimization.

\subsection{Virtual Constraint based Feedback Controller}
\label{sec:virt-constraint}
des
In this paper, we focus on the periodic walking of the exoskeleton at a desired forward velocity. Let $\pi$ be a set of parameters that describe the desired behaviors; the virtual constraints are 
the difference between the actual outputs, ($\mathbf{y}^a(\mathbf{q})\in \R^{12}$), and the desired outputs, ($\mathbf{y}^d(\tau, \alpha_\pi) \in \R^{12}$), defined as 
\begin{align}
  \mathbf{y}(\mathbf{q}, \tau, \alpha_{\pi}) := \mathbf{y}^a(\mathbf{q}) - \mathbf{y}^d(\tau, \alpha_{\pi}),
  \label{eq:vc}
\end{align}
where, $\tau = \frac{t-t_0}{T} \in [0,1]$ is the scaled parameterization of time for each domain with $t_0$ being the time at the beginning of the domain, $T$ is the duration of the domain, 
and $\mathbf{q}$ is the set of actual joint angles. Here,  $\alpha_{\pi}$ is a set of coefficients of B\'{e}zier  polynomials and we are using it to represent the desired outputs here. The subscript indicates that these coefficients are determined by the desired behaviors $\pi$, typically through the gait optimization discussed in the previous subsection. The actual outputs, $\mathbf{y}^a(\mathbf{q})$, can be chosen as some kinematic functions, such as leg length or leg angle, to be directly regulated by the controller. In this paper, we particularly choose the actuated joint angles as our outputs for simplicity, i.e., $\mathbf{y}^a(\mathbf{q}) = \mathbf{q}$. 

Given a set of virtual constraints, one can construct a feedback tracking controller to drive the actual outputs to the desired outputs, i.e., $\mathbf{y}(\mathbf{q},\alpha_{\pi}) \to 0$. Considering the significant model uncertainty that exists in the human-exoskeleton system, we use a simple model-independent Proportional-Derivative(PD) control scheme, given as
\begin{align}
  u = -K_p(\mathbf{q}^a - \mathbf{q}^d) - K_d(\mathbf{\dot{q}}^a - \mathbf{\dot{q}}^d),
  \label{eq:PD}
\end{align}
where $K_p$ and $K_d$ are the proportional and derivative control gain matrices, respectively. As discussed in \secref{section3}, the adaptability of our proposed approach will be through modifying the desired outputs, $\mathbf{y}^d$, via the neural-network based feedback regulator to improve the stability and behavior tracking performance. 

\section{NN based Adaptive Controller} \label{section3}
In this section, we propose an adaptive feedback regulator using a NN to realize stable walking while regulating forward walking speed in the presence of model and environmental uncertainties.


\subsection{Review of Feedback Regulators}
\begin{figure}
\centering
\vspace{2mm}
\includegraphics[trim={0cm 0cm 0cm 0cm},clip,width=1\columnwidth]{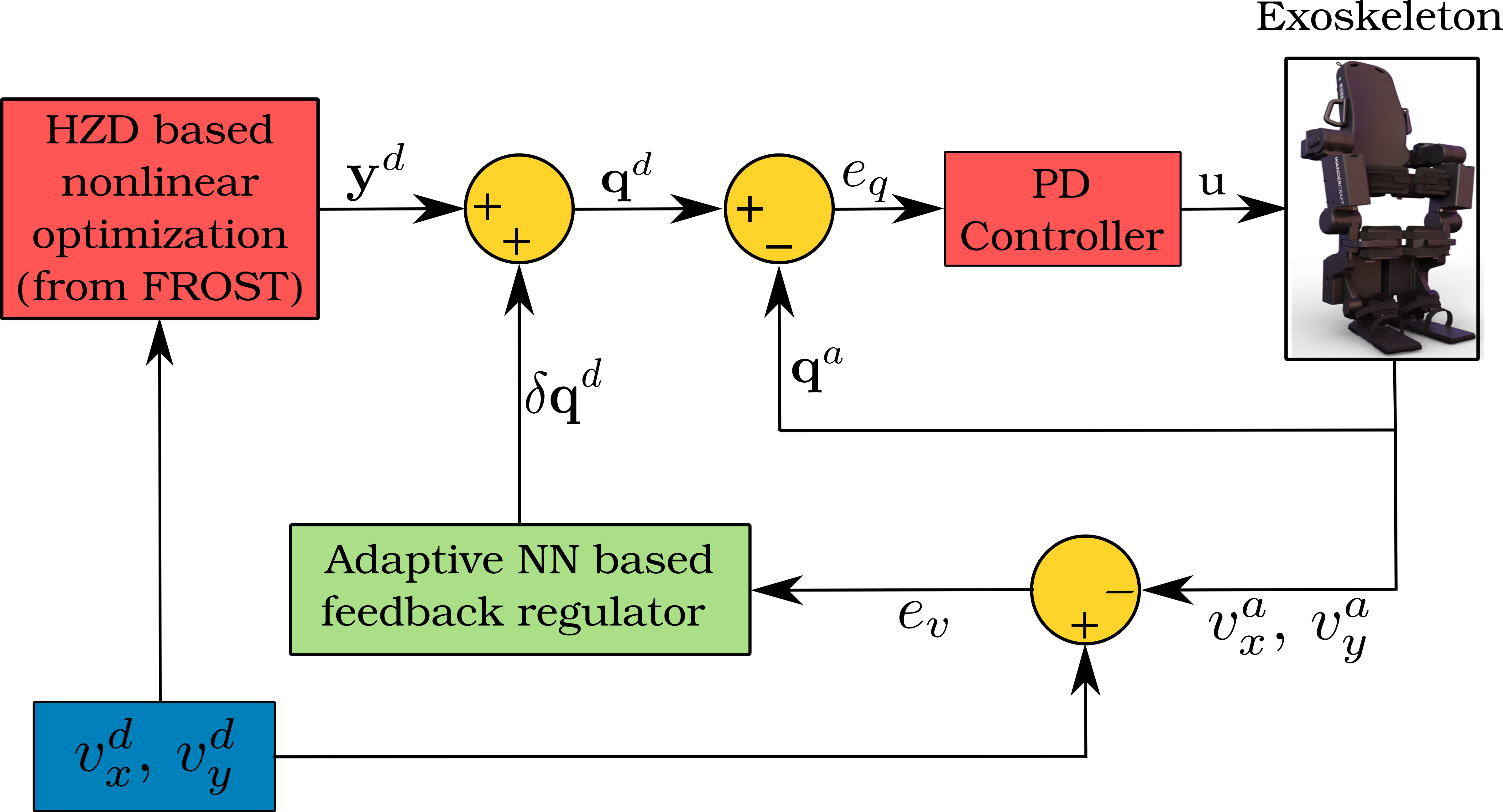}
\caption{The feedback structure of the adaptive regulator (green box) integrated into the virtual constraint based feedback controller. } 
\label{fig:structure}
\end{figure}

The presence of uncertainties in the dynamical model of a robot can generate an unstable walking motion. This leads us to the primary motivation behind the regulator design, which is essentially modifying the desired optimized trajectories to suit the real model better and enhance its stability. However, the complex dynamics of 3D bipedal robots render the regulator design improving the stability of the walking motion highly challenging.  To address this challenge, researchers have started employing heuristically-designed feedback regulators (see ~\cite{Kolathaya2014Composing, Gong2019Feedback}) in addition to the virtual constraint based feedback controller (see ~\secref{sec:virt-constraint}). ~\figref{fig:structure} illustrates how the virtual constraint based feedback controller and the feedback regulator work in conjunction.  

\begin{figure}
\centering
\vspace{2mm}
\includegraphics[trim={0cm 0cm 0cm 0cm},clip,width=0.70\columnwidth]{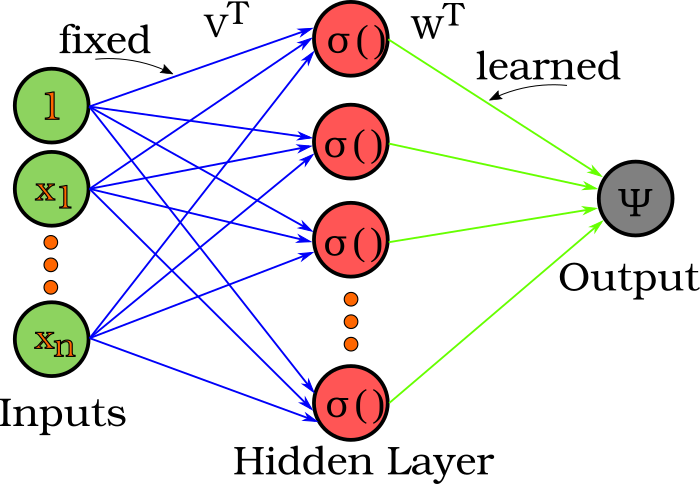}
\caption{The structure of a single-layer neural network policy. The weights of the first layer are randomly initialized and are kept fixed. The second layer weights are learned with the delta-rule developed in \secref{section3}. } 
\label{fig:NN}
\end{figure}

Foot placement is a widely used heuristically-designed regulator as highlighted in ~\cite{Da2016From2D, Rezazadeh2015Spring, Gong2019Feedback}. In particular, this regulator often uses decoupled structures, relating certain joints with specific desired features, which in this case are forward and lateral velocities. For instance, this regulator adds an offset to desired trajectories of the swing hip sagittal joint and the swing hip frontal joint (see \figref{fig:atlante}) to regulate forward and lateral velocities, respectively. This modification in the desired trajectories will reshape the virtual constraints since we are choosing $\mathbf{y}^a(q) = \mathbf{q}$ as explained in the ~\secref{sec:virt-constraint}. Modified virtual constraints can be expressed as  
\begin{align}
  \mathbf{y}(\mathbf{q}, \tau, \alpha_{\pi}) &:= \mathbf{q}^a - (\mathbf{q}^d(\tau, \alpha_{\pi}) + \delta \mathbf{q}^d),
  \label{eq_vc_deltay}
\end{align}
where, $\delta \mathbf{q}^d$ represents the trajectory compensation in the desired output $\mathbf{y}^d$, which is equal to the offset provided by a feedback regulator. Instead of producing an offset for all the desired joints angles, we focus on the most relevant ones in the regulation of longitudinal and lateral speed: swing sagittal hip and swing frontal hip, respectively, represented by ${\delta \mathbf{q}}^d_{x}[k]$ and ${\delta \mathbf{q}}^d_{y}[k]$.
The formulation of this feedback regulator resembles the conventional PD controller, as expressed below.
\begin{equation}
    \begin{aligned}
        \label{eq_deltay_est}
        {\delta \mathbf{q}}^d_{x}[k] &= K_{p_{x}}(v^{d}_{x} - v_{x}[k]) + K_{d_{x}}(v_{x}[k]-v_{x}[k-1]), \\
        {\delta \mathbf{q}}^d_{y}[k] &= K_{p_{y}}(v^{d}_{y} - v_{y}[k]) + K_{d_{y}}(v_{y}[k]-v_{y}[k-1])\\
    \end{aligned}
\end{equation}  
where, $v_{x}[k]$ and $v_{y}[k]$ are the average longitudinal and lateral speeds of the robot,  $v^{d}_{x}$ and $v^{d}_{y}$ are the reference speeds and, $K_{p_{x}}, K_{d_{x}}, K_{p_{y}}, K_{d_{y}}$ are manually tuned gains. However, the major shortcoming of this regulator is that it can not effectively address large model uncertainties (e.g., changes in weight and its distribution) and external disturbances. For ATALANTE, the standard weight of a patient is considered to be 60kg, and the maximum allowed weight is 90kg. Considering the high level of variability, it is indispensable to have a regulator that can learn to compensate the unknown system dynamics without prior knowledge of it and address the issue of stability and forward velocity tracking. This brings us to the main contribution of this study, the adaptive feedback regulator, which is covered in detail in the following sections.

\subsection{Adaptive Feedback Regulator} \label{adapt_reg}
\begin{figure}
\centering
\vspace{2mm}
\includegraphics[trim={0cm 0cm 0cm 0cm},clip,width=0.9\columnwidth]{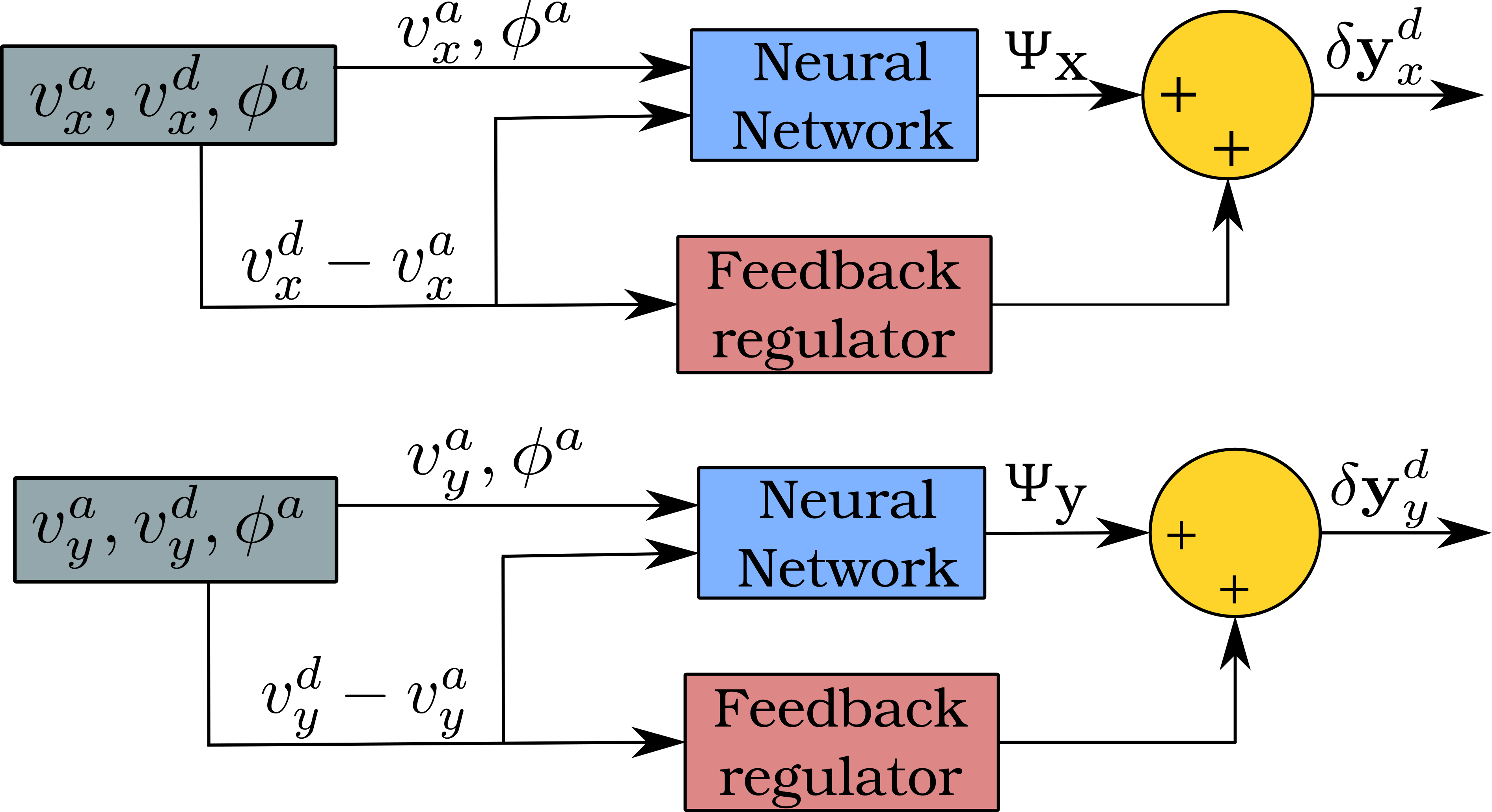}
\caption{Detailed structure of the proposed adaptive feedback regulator} 
\label{fig:struct_adap_control}
\end{figure}

In this subsection, we propose an adaptive feedback regulator to remove the limitations of heuristically designed feedback regulators. Here, we describe the underlying idea that brought us to this specific structure of the regulator. Like a heuristically designed regulator, we aim to modify certain joint trajectories to regulate the forward and lateral speeds. Let $v[k]$ be an average velocity at instant $k$ and its desired value $v^d$ (we are omitting the subscript $x$ or $y$ since the idea is the same for both longitudinal and lateral speed). When the robot's foot is in contact with the ground, we can express its actual velocity with the jacobian as $v[k] = J(q[k]) \dot{q}[k]$. To produce a change in the velocity, we can modify the joint angles by $\delta q$. In the ideal case, with an adequate $\delta q$ we aim that actual speed matches to $v^d$, i.e 
\begin{align}
        v^d&= J(q[k] + \delta q[k]) \dot{q}[k] \\
        v^d &= \underbrace{J(q[k])\dot{q}[k]}_{v[k]} + \delta q[k] \frac{\partial J(q[k])}{\partial q} \dot{q}[k] + \underbrace{\mathbb{O}(\delta q, q[k], \dot{q}[k])}_{h.o.t} \\
        \delta q[k] &= P(\delta q, q, \dot{q}) (v^d - v[k]) + Q(\delta q, q, \dot{q})
\end{align}
where, $P$ and $Q$ are not-computed, unknown functions of $(\delta q, q, \dot{q})$. This provides the idea of adding $\Psi_x$ and $\Psi_y$ in our existing feedback regulators to compensate for those functions, which brings us to the foundation of the adaptive feedback regulator. Let $v_{x}[k]$ and $v_{y}[k]$ be the average longitudinal and lateral speeds of the robot in the middle of a step $k$, $v^{d}_{x}$ and $v^{d}_{y}$ be the reference speeds, we define







\begin{equation}
    \begin{aligned}
        \label{eq_deltay_est}
        {\delta \mathbf{q}}^d_{x}[k] &= K_{p_{x}}(v^{d}_{x} - v_{x}[k]) + K_{d_{x}}(v_{x}[k]-v_{x}[k-1]) + \Psi_{x}, \\
        {\delta \mathbf{q}}^d_{y}[k] &= K_{p_{y}}(v^{d}_{y} - v_{y}[k]) + K_{d_{y}}(v_{y}[k]-v_{y}[k-1]) + \Psi_{y}, \\
    \end{aligned}
\end{equation}    
where, ${\delta \mathbf{q}}^d_{x}$ will be added as an offset to desired swing sagittal hip and ${\delta \mathbf{q}}^d_{y}$ will be added to desired swing frontal hip. A detailed structure for the decoupled controllers is presented in \figref{fig:struct_adap_control}.

Since our primary goal is to compensate for the unknown functions$P(\delta q, q, \dot{q})$ and $Q(\delta q, q, \dot{q})$ with added terms $\Psi_x$ and $\Psi_y$, provided that the controller needs to learn them online, we refer to the widely known approaches in the neural networks community to use them as a nonlinear function approximators to learn un-modeled system dynamics. Motivated by the work of ~\cite{Narendra1997Adaptive, Sanne1995Stable, Chen1992Adaptive, DeWolf2016, Gnucci2019OnTheAdaptive, Lewis1998Neural}  we use a single layer NN shown in Fig.~\ref{fig:NN} as a nonlinear function approximator without the need for computing a regression matrix. The significant advantage of NN-based controllers is that they can virtually approximate any smooth function with an appropriate number of neurons and layers. 
\subsection{Adaptive  Structure}
In this subsection we focus on the construction of the structure reflecting the terms $\Psi_{x}$ and $\Psi_{y}$ highlighted in equation \eqref{eq_deltay_est}. These are the outputs of the individual NNs multiplied with gains as expressed below:
\begin{equation}
    \begin{aligned}
    \label{psi}
    \Psi_i = \mathbf{k_i}\hat{\mathbf{W_i}}^\intercal \sigma (\hat{\mathbf{V_i}}^\intercal \mathbf{x}),  \ (i = x,y)
    \end{aligned}
\end{equation}
where, $\mathbf{k_x}$ and $\mathbf{k_y}$ are the gains of the adaptive learning term, $\sigma$ is the activation function, $\hat{{\mathbf{W}}}$ and $\hat{{\mathbf{V}}}$ are output and input layer weight matrices respectively, and $\mathbf{x}$ is the input vector. Moreover, we have chosen the Sigmoid function as an activation function for the purpose.  As shown in the \figref{fig:struct_adap_control} inputs for the NN corresponding to $\Psi_x$ are actual forward velocity, torso orientation, and the error between desired and actual forward velocity. Similarly inputs for the NN corresponding to $\Psi_y$ can be interpreted from the \figref{fig:struct_adap_control}. As highlighted in \cite{Lewis1998Neural}, we are normalizing the input vector and then appending the input array by an additional input unity before feeding it to the Sigmoid function. Each of the structures has only one hidden layer with a thousand neurons.

 For simplicity, we are not learning the input layer weights $\hat{V}$. These have been initialized with normal distribution until we found values good enough for the approximation and kept constant throughout the simulation process. $\hat{W}$ are the output layer weights, and we are learning them via gradient descent delta-rule in an online manner. The weight update rule for the given structure can be expressed as, 
\begin{align}
    \label{weight_update}
    \Delta w_{i,j} = -\gamma E_j h_i
\end{align}
where, $w_{i,j}$ is the weight from the $i^{th}$ hidden neuron to the $j^{th}$
output, $E_j$ is the error signal for the $j^{th}$ output, $h_i$ is the output of
the $i^{th}$ hidden neuron, and $\gamma$ is the learning rate chosen as
$\gamma = 1e{-4}$. However, in the current scenario, we have only one output for each of the structures rendering $\Delta w$ to be a vector. We have chosen the difference between desired and average actual velocity as an error signal as we aimed to regulate it.

\section{SIMULATION RESULTS} \label{section4}
The proposed method is validated in a dynamics simulation of ATALANTE using the Jiminy Simulator (see ~\cite{jiminy}). An Invariant extended Kalman filter (InEKF) ~\cite{Ross2020EKF} has been employed to estimate the orientation of the torso and velocity of the robot from the data generated via IMU available on the torso of the exoskeleton. A single walking period consists of two single support and two double support phases. A walking gait without a double support phase could be detrimental to stability. Since its high inertia, exoskeleton, plus patient system can lead to additional external forces while switching to and from single support domains.

In this section, we present the results of the adaptive feedback regulator when the actual robot model has different mechanical properties such as mass variation and different mass distribution than the one used to generate the gait. We also present the regulator performance for different walking speeds of the robot. In addition, we show the robustness of the regulator via applying external impulsive forces on the robot during the walking gait. In the end, we conclude this section by providing a comparative study between the traditional feedback regulator and the novel NN based adaptive regulator. 
In all the simulation (both adaptive and traditional regulator) runs, we are starting the forward foot placement regulator at 20 seconds as the robot needs around 10 seconds to reach a steady-state walking after starting from rest. However, the lateral foot placement regulator is active from the beginning of the walking motion to maintain stability in the presence of uncertainties. This also explains the slight difference in steady-state for the adaptive regulator and the conventional feedback regulator. 

\begin{figure}
\centering
\vspace{1mm}
\includegraphics[trim={0cm 0cm 0cm 0cm},clip,width=0.9\columnwidth]{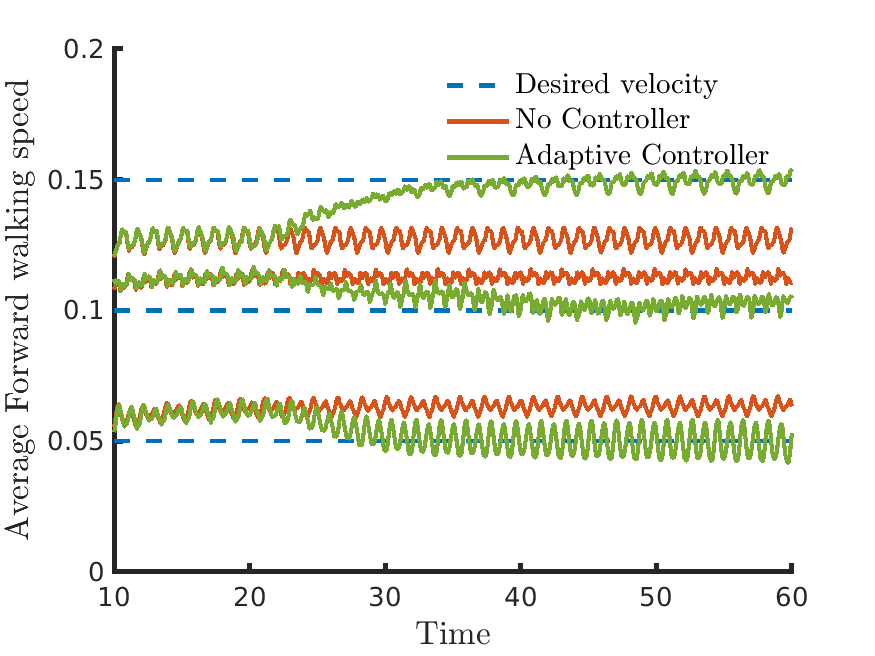}
\caption{Controller Performance for forward walking speed of $0.05m/s$, $0.1m/s$ and $0.15m/s$} 
\label{diff_vel}
\end{figure}

\subsection{Performance for different forward speeds}
First, we validated the efficacy of the regulator by tracking different forward walking speeds. \figref{diff_vel} demonstrates the performance of the adaptive regulator for walking speeds $0.05m/s , 0.1m/s$ and $0.15m/s$. It is interesting to observe in the same figure that the robot can not catch up with the desired forward speed when none of the controllers is employed despite the optimized trajectories are designed for that specific speed. This deviation can be attributed to the assumptions we have made while setting up the optimization problem. For instance, rigid ground impact of the foot, absence of the friction and damping in motored joints, and varied friction coefficient between ground and contact points of the foot to name a few. Moreover, it is difficult to render the error between desired and actual trajectories to absolute zero regardless of the choice of controller. This mismatch in the actual and desired trajectories can lead to a bundle of new challenges such as untimely and undesired orientation ground impact of the foot and more. These factors can work against the desired forward speed and overall stability of the robot. 

\begin{figure}
\centering
\vspace{1mm}
\includegraphics[trim={0cm 0cm 0cm 0cm},clip,width=0.9\columnwidth]{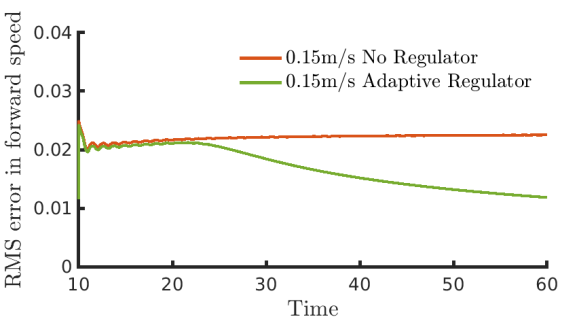}
\caption{RMS error in forward walking speed for 0.15m/s} 
 \label{rms_error}
\end{figure}

Further, to highlight the regulator's capability, we present the RMS error plot (\figref{rms_error}) of the forward velocity for $0.15m/s$. Error is evolving quite similarly until 20 seconds for both the cases, one without the controller and the other in the presence of the adaptive controller since the controller is inactive until then. But, after that controller takes few seconds to learn the parameters of the NN and drives RMS error plot downwards to minimize it. 


\subsection{Response to changes in the model properties of the robot}
\begin{figure}
\centering
\vspace{1mm}
\includegraphics[trim={0cm 0cm 0cm 0cm},clip,width=0.80\columnwidth]{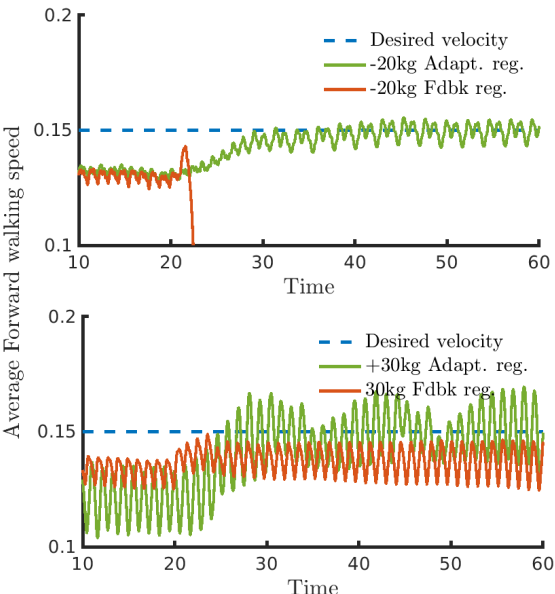} 
\caption{Response of the novel adaptive and the conventional feedback regulator when mass of the pelvis link changed} 
\label{mass_change}
\end{figure}

The central theme of this section is to demonstrate the efficacy of the regulator by maintaining stability and tracking forward velocity regardless of model uncertainties. For instance, variation in mass and mass distribution. We changed the model mass by varying the pelvis link (corresponding to the heaviest link) mass from its standard value in the URDF file. This is to simulate the controller performance for patients with different body weights. The gait has been generated considering a total system mass of 135kg (75kg exoskeleton mass and 60kg of mannequin mass). The maximum allowed patient mass on the exoskeleton is 90kg. Therefore, we show the regulator performance in ~\figref{mass_change} by increasing pelvis mass up to 30kg and reducing it up to 20kg.

\begin{figure}
\centering
\vspace{1mm}
\includegraphics[trim={0cm 0cm 0cm 0cm},clip,width=0.8\columnwidth]{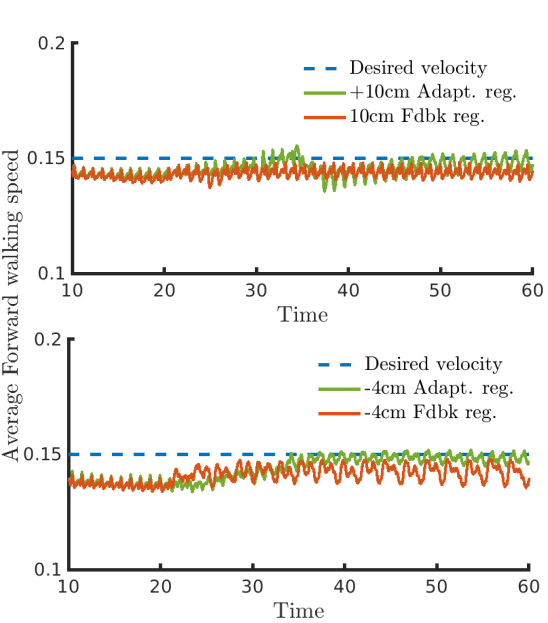} 
\caption{Response of the novel adaptive and the conventional feedback regulator when center of mass location of the pelvis link changed in longitudinal direction} 
\label{com_change}
\end{figure}

The fundamental idea behind the powered lower-limb exoskeleton is to achieve hands-free walking. This feature can allow the user to engage the upper body in several different activities; for instance, the user carrying objects of different weights and shapes. This idea can be translated as the uncertainty in mass distribution in the simulation model.  We achieve that by adding an offset to the COM location in the URDF file. We are adding an offset of a maximum $0.1m$ in forward and $0.04m$ in the backward direction. From the ~\figref{com_change} we can verify that regulator successfully compensates for unknown dynamics and leads stable walking, ensuring convergence to desired forward velocity.   

\begin{figure}
\centering
\vspace{1mm}
\includegraphics[trim={0cm 0cm 0cm 0cm},clip,width=0.85\columnwidth]{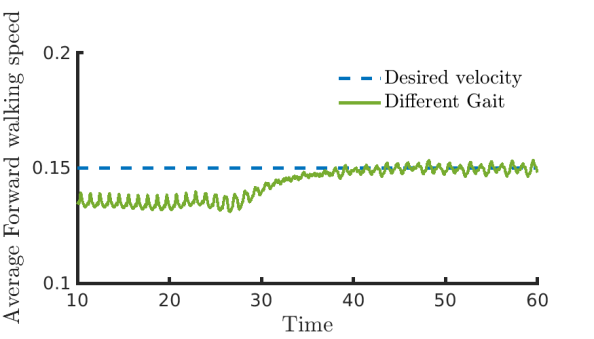} 
\caption{Controller performance for the gait generated using different model than the one in simulation}
\label{diff_gait}
\end{figure}

The generation of a walking gait is heavily dependent on the mechanical properties of the patient and exoskeleton system and it is evident from \eqref{eq:gait-opt-frost}. This makes it very cumbersome to repeat the whole process for each patient with a different physical build on a larger scale. Therefore, we show the adaptability of the presented regulator by generating the walking pattern for a different model than the one in the simulation.  ~\figref{diff_gait} illustrates the performance of the regulator, which leads the steady velocity achieved at 20 seconds to the desired velocity of the gait.

\subsection{Robustness}
\begin{figure}[h]
\centering
\vspace{1mm}
\includegraphics[trim={0cm 0cm 0cm 0cm},clip,width=0.8\columnwidth]{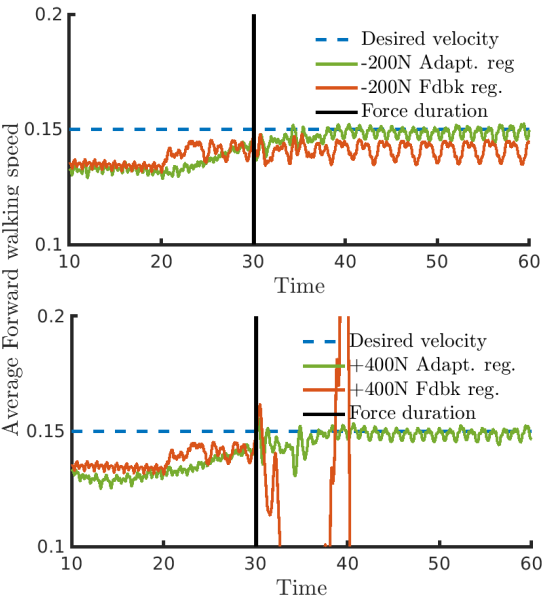}
\caption{Performance of the novel adaptive and the conventional feedback regulator when forward and backward impulsive forces are applied} 
\label{adversarial}
\end{figure}

We demonstrate the robustness of the regulator by validating its capability to recover the forward speed under external disturbances. For this purpose, we have applied both the forward force of $400 N$ and the backward force of $200 N$ for 0.1sec in separate simulations after $30 sec$ from the start of the simulation. ~\figref{adversarial} shows that the regulator can handle external forces successfully without falling and, more importantly, recovering the tracking of the forward velocity after the disturbance is applied.

\subsection{Comparison between Heuristic and Adaptive controller}
\label{sec:comparison}

To illustrate the impact of the adaptive regulator, we are also presenting a comparative study against the traditional feedback regulator. ~\figref{mass_change} shows the forward velocity tracking performance of both the regulators for a walking speed of $0.15m/s$. When the pelvis link's mass was reduced by $20 kg$ traditional feedback regulator failed to maintain the stability, let alone the tracking performance. Similar behavior can be observed for the feedback regulator from the ~\figref{adversarial} when $200N$ of impulsive force applied in the backward direction. Although, for the traditional regulator case robot did not fall when the  location of COM changed (see ~\figref{com_change}) but walking motion develops a steady state error in velocity tracking. In summary, either the traditional regulator fails to balance the robot or develops steady state  tracking error when pushed to the limits. On the contrary, the novel adaptive feedback regulator shows consistent performance regardless of the uncertainty put forth.




\section{Conclusion}
\label{sec:conclusion}
In this paper, we propose the addition of two double support domains in a nominal walking gait. We also present the design of the adaptive regulator leveraging the idea of NN for nonlinear function approximation and evaluate its performance over the ATALANTE platform. The regulator assures the tracking of the forward velocity of an exoskeleton under the presence of model and environmental uncertainties. The performance has been demonstrated using the Jiminy simulator. The presented regulator builds upon the existing ideas of the feedback regulator to handle the uncertainties more effectively. It gradually learns and converges at the modifications for the specific joint trajectories. The regulator's online learning nature makes it even more effective to learn unknown system dynamics without any prior training. This regulator has excellent performance on the simulator. However, it lacks the experimental performance evaluation, which is the next target for this research. Moreover, the regulator has been evaluated particularly for an exoskeleton but being able to learn system dynamics, it can be implemented for any bipedal walking robot for stability and forward velocity tracking.   

\bibliography{bib/references.bib}
\bibliographystyle{IEEEtran}

\end{document}